\documentclass[oldversion]{aa}

\usepackage{natbib}  

\usepackage{graphicx}
\usepackage{txfonts}

\def\hi{H{\sc i\,}}
\def\farcs{\hbox{$.\!\!^{\prime\prime}$}}
\begin{document}
\bibliographystyle{aa} 
   \title{Stars and gas in the very large interacting galaxy NGC~6872}
   \author{C. Horellou
          \inst{1}
          \and
          B. Koribalski
	\inst{2}
          }

   \offprints{C. Horellou}

   \institute{Onsala Space Observatory, 
        Chalmers University of Technology, 
        SE-439 92 Onsala, Sweden \\
              \email{Cathy.Horellou@chalmers.se}
         \and
Australia
Telescope National Facility, CSIRO, P.O. Box 76, Epping, NSW 1710, Australia\\
             \email{Baerbel.Koribalski@csiro.au}
             }

   \date{Received 13 July 2006 / Accepted 2 November 2006}

   \abstract{
The dynamical evolution of the  
large ($>100$~kpc), barred spiral galaxy NGC~6872 
and its small companion IC~4970 in the southern group Pavo
is investigated.  
We present 
$N$-body simulations with stars and gas and 
21~cm \hi observations carried out with the Australia Telescope Compact Array 
of the large-scale distribution and kinematics of atomic gas. 
\hi is detected toward the companion, corresponding to a gas mass of 
$\sim 1.3\times 10^9 M_\odot$. 
NGC~6872 contains $\sim 1.4\times 10^{10} M_\odot$ of \hi gas, 
distributed in an extended rotating disk.
Massive concentrations of gas ($\sim 10^9 M_\odot$)
are found at the tip of both tidal 
tails and towards the break seen in the optical northern arm near
the companion. 
We detect no \hi counterpart to the X-ray trail
between NGC~6872 and NGC~6876, the dominant elliptical galaxy
in the Pavo group located $\sim 8'$ to the southeast.
At the sensitivity and the resolution of the observations,
there is no sign in the overall \hi distribution that
NGC~6876 has affected the evolution of NGC~6872.
There is no evidence of ram pressure stripping either.
The X-ray trail could be due to gravitational focusing of 
the hot gas in the Pavo group 
behind NGC~6872 as the galaxy moves supersonically through the
hot medium. 
The simulations of a gravitational interaction with a small
nearby companion on a low-inclination prograde passage
are able to reproduce most of the
observed features of NGC~6872, including the
general morphology of the galaxy, the inner bar,
the extent of the tidal tails and
the thinness of the southern tail. }


\authorrunning{Horellou \& Koribalski}
\titlerunning{Stars and gas in NGC~6872/IC~4970}

   \maketitle

\section{Introduction}
\begin{table*}[t]
\def\hf{\hfill}
\halign
{#&#&#&#&#\cr
\hline
\noalign{\smallskip}
\hf		&Note	& NGC~6872\hf 	&IC~4970\hf			&NGC~6876\hf\cr
\noalign{\smallskip}
\hline
\noalign{\smallskip}
Other names\hf	&\hf	& ESO~073-IG~32\hf	& ESO~073-IG~33		&ESO~073-IG035\cr
\hf		&\hf	& VV~297a\hf		&VV~297b\hf\cr
\hf		&	& AM~2011-705\hf\cr
$\alpha$(J2000) &(1)	& $20^{\rm h}16^{\rm m}56\fs91$	&$20^{\rm h}16^{\rm m}57\fs87$ 	&$20^{\rm h}18^{\rm m}19\fs15$\cr
$\delta$(J2000) &(1)	&$-70^\circ 46'04\farcs5$ &$-70^\circ 44' 57\farcs3$ 		&$-70^\circ 51'31\farcs7$\cr
Type		&(2)	&.SBS3P.  	&.LA.-P*					&SB0\cr
Systemic velocity &(2)	&4818~km s$^{-1}$\hf		&4727 km s$^{-1}$\hf		&4010 km s$^{-1}$\hf\cr
Distance 	&(3)	& 61 Mpc\hf\cr
Scale		&	& 17.75 kpc arcmin$^{-1}$\cr
		&	& 296 pc arcsec$^{-1}$\hf\cr
$D_{25}$	&(2)	& 6$\farcm$025=106.9 kpc\hf	&0$\farcm$676=12 kpc\hf		&2$\farcm$8\cr
Position angle\hf&(2)	& 66$^{\circ}$			&6$^\circ$			&79$^\circ$\cr
Axis ratio\hf 	&(2)	& 0.288=cos(73$^{\circ}$)       &0.323=cos(71$^\circ$)		&0.786=cos(38$^\circ$)\cr
$f_{60 \mu{\rm m}}$	&(4)	&1.67 Jy\hf\cr
$f_{100 \mu{\rm m}}$	&(4)	&6.61 Jy \hf\cr
$T_{\rm FIR}$		&(4)	&28 K\hf\cr
$L_{\rm FIR}$		&(4) 	&$1.6\times10^{10} L_{\odot}$\hf\cr
$L_B$		&(2)	& $1.67\times10^{11}L_\odot$\hf	&$1.24\times10^{10} L_\odot$\hf 	&$1.07\times 10^{11} L_\odot$\hf\cr
H{\sc i} mass\hf&(5)	& $1.41\times10^{10} M_\odot$\hf	&$1.3\times 10^9 M_\odot$\hf\cr
H$_2$ mass\hf	&(6)	& $9.6\times10^{8} M_\odot$\hf\cr
\noalign{\smallskip}
\hline
}
\smallskip
Notes: (1) Position of NGC~6872: \cite{1990AJ.....99.2059R}, 
uncertainty of 2$\farcs$5. 
Position of IC~4970: \cite{1982euse.book.....L}, 
uncertainty of 7$\farcs$5.  
Position of NGC~6876: \cite{2006AJ....131.1163S}, 
uncertainty of 1$\farcs$25.
(2) Systemic velocity of NGC~6876: \cite{1995A&AS..110...19D}; 
other data: \cite{1991trcb.book.....D}. The blue-band luminosities have been 
derived from the blue-band magnitudes corrected for extinction, $B_T^0$,
using the following relation: 
$\log(L_B/L_\odot)= 12.192 -0.4 B_T^0 + \log(D/{\rm 1 Mpc})^2$, 
consistent with the blue magnitude for the Sun $M_B^0= 5.48$. \\ 
(3) Consistent with $H_0$ = 75 km s$^{-1}$ Mpc$^{-1}$. \\
(4) \cite{1990IRASF.C......0M}. 
The far-infrared luminosity $L_{FIR}$, in solar luminosities, 
has been derived from the flux densities at 60 and 100 $\mu$m listed in the IRAS 
catalogue using the relation:
$L_{FIR}= 3.94\times10^5 D^2 (2.58 f_{60} + f_{100})$
where $D$ is the distance in Mpc, $f_{60}$ and $f_{100}$ are the flux densities in Jy. 
The dust temperature $T_{FIR}$ has been derived from $f_{60}$ and $f_{100}$ 
by fitting a black body and assuming that the dust emissivity is
inversely proportional to the wavelength. 
\\ 
(5) This work.
\\
(6) \cite{1997A&AS..126....3H}. 
The H$_2$ mass refers to the central 
45$''$ of NGC~6872 and was derived from CO(1-0) observations 
using a standard CO-H$_2$ conversion factor. 
\caption[]{Basic parameters of NGC~6872/IC~4970.}
\end{table*}

Galaxies with extended tidal tails form an outstanding laboratory to 
study the effect of a collision on the different components of a galaxy 
because of the sensitivity of the morphology and kinematics of the tails  
to the initial distribution of the matter (both luminous and dark)
and to the geometry of the interaction 
(e.g., 
\citealt{1999ApJ...526..607D},   
\citealt{1999MNRAS.307..162S}). 
Early simulations clearly showed that  
prograde, co-planar encounters are more efficient 
than retrograde encounters 
at triggering long thin tails
(e.g., 
\citealt {1941ApJ....94..385H}, 
\citealt{1961ZA.....51..201P},  
\citealt{1963ZA.....58...12P},  
\citealt{1972ApJ...178..623T}). 
The tails are made of stars and gas 
that have been thrown out 
from the galactic disks during the gravitational interaction 
which may lead to the merger of the two disks. 
Spectacular examples of nearby galaxies at different stages of 
an interaction can be found in the Arp atlas 
(\citeyear{1966ApJS...14....1A}) 
and the Arp \& Madore catalog 
(\citeyear{1987arpmadore}). 
The tails are often gas-rich and their kinematics can be traced out to 
large radii by 
21~cm line observations of atomic hydrogen 
(e.g., \citealt{1996AJ....111..655H}). 
Some tidal tails contain 
massive self-gravitating concentrations of matter which 
may be the progenitors of the so-called tidal dwarf galaxies
(\citealt{1993ApJ...412...90E}, 
\citealt{2003A&A...411L.469B},  
\citealt{2004A&A...425..813B}). 
Tidal tails develop not only in merging systems, but also in 
flyby encounters, where the interacting galaxies do not form a bound system. 
It is of particular interest to study tidal tails at an early time of 
the interaction when the two galaxies are well separated and have a clear 
morphology. 
A few 10$^7$~yr after closest approach, the tails are already well developed
and it is possible to reconstruct the dynamical history of the interaction
and examine its influence on the different components of a galaxy 
and the level of induced star formation. 

The southern galaxy NGC~6872 
is one of the largest spiral galaxies known. 
Star formation is traced all along the arms, which extend over
more than 100~kpc at our adopted distance of 61~Mpc. 
The galaxy belongs to a small group, Pavo (see Fig.~1a). 
NGC~6872 is likely to be affected by tidal perturbations 
from the nearby companion IC~4970, a small lenticular galaxy 
located 1$\farcm$1 to the north. 
\cite{2005ApJ...630..280M} 
have discovered a more than 100~kpc long X-ray trail extending between NGC~6872
and the neighbor galaxy NGC~6876, the dominant elliptical in the group
which lies about 8$'$ ($\sim 142$~kpc) to the southeast. 
The radial velocity of NGC~6876 is about 800~km~s$^{-1}$ 
lower than that of NGC~6872 (see Table~1). 
The X-ray trail is hotter ($\sim 1$ keV) than the undisturbed Pavo intergalactic medium 
($\sim 0.5$ keV) and has a low metal abundance. 
The authors interpret the trail as partly due to gravitational focusing 
of the intracluster gas into a Bondi-Hoyle wake, as the spiral galaxy moves
supersonically through the intracluster medium; 
they point out that 
the trail could also consist of a mixture of intracluster gas 
and gas removed from NGC~6872 by turbulent viscous stripping.  

The spectacular Very Large Telescope (VLT) multicolor image 
of NGC~6872/IC~4970 
displays striking contrasts between the inner region,
especially the straight northern arm,
and the blue diffuse tidal tail to the north-east 
(ESO Press Release 20b/99; see also Fig.~1b, 
which shows the VLT blue-band image on a grey scale). 
A bar is clearly seen in the 2MASS $J, H, K$
images \citep{2003AJ....125..525J}. 
Fabry-P\'erot H$\alpha$ observations have revealed the presence of
ionized gas at the tip of both tails
\citep{1993ApJ...418...82M}. 
The location of the ionized regions coincides
with that of young, blue
stellar clusters seen in the optical images.
No H$\alpha$ emission was detected from the central region and the
bar, nor 
from the early-type companion IC~4970. 

\cite{2005A&A...435...65B} 
have studied the rich population of star clusters in NGC~6872 
using archival VLT images in the $B$, $V$, and $I$ bands, complemented 
with new VLT observations in the $U$-band. 
They estimated the mass, age and extinction of the star clusters 
and found that most of the young massive clusters 
(of mass between 10$^4$ and 10$^7$ M$_\odot$ and 
less than 100~Myr old) are located 
in the tidal tails or in the outer part of the galactic disk. 
The mass distribution of the star clusters follows a power-law 
with an index similar to that found in other galaxies. 
The authors also estimated the star formation rate in different regions of the
galaxy and showed that the northern tail is forming stars at about twice
the rate of the southern tail, and about five times the rate of the
main body. 

\cite{1993ApJ...418...82M} have modeled the dynamic evolution of NGC~6872, 
as it is gravitationally perturbed
during the passage of the small companion IC~4970 on a low-inclination, prograde orbit. 
They were able to reproduce most of the observed features, although 
the simulated tidal tails were not as thin as the observed ones. 
The model predicted 
an accumulation of gas toward the center, which was subsequently 
detected in CO emission (\citealt{1997A&AS..126....3H}). 

In this paper, we present new maps of the large-scale
distribution and kinematics of atomic hydrogen (\hi) in the NGC~6872/IC~4970 system 
and $N$-body simulations with stars and gas.  
With a total \hi mass $> 1.5 \times10^{10} M_\odot$
(\citealt{1997A&AS..126....3H}, 
\citeyear{1999Ap&SS.269..629H}), 
the pair is among the most gas-rich galaxies in the 
optically selected sample 
of $\sim60$ interacting and merging galaxies 
compiled by \cite{1990A&AS...86..167J}. 
The single-dish \hi spectrum is extremely broad 
($\sim950$ km~s$^{-1}$ at the base of the \hi line) and observations at a resolution higher 
than the Parkes 15$'$ beam were required to determine the location of the 
\hi gas. 
The Australia Telescope Compact Array (ATCA) 
in several configurations 
makes it possible to obtain data 
with angular resolution ($\sim40''$, corresponding to about $\sim12$~kpc 
at our adopted distance) and a sensitivity 
adequate to detect extended \hi emission. 
We pointed the array 
half-way between the large spiral galaxy and NGC~6876, 
the dominant elliptical in the Pavo group 
located ca 8$'$ to the south-east, 
in order to be more sensitive to diffuse \hi gas in the Pavo group 
and detect traces 
of a possible gravitational interaction 
in the extended \hi distribution. 

We also discuss $N$-body simulations with stars and gas performed using 
a particle-mesh code, where the gas is modeled as a collection of clouds
that dissipate energy through inelastic collisions. 
Preliminary results have been presented elsewhere 
(\citealt{2003Ap&SS.284..499H}, \citealt{2004IAUS..217..422H}). 
We have used simulations to investigate the evolution of gas and stars 
in such a close encounter,  
examine the influence of the dark matter halo 
and of the geometry of the collision on
the characteristics of the tidal tails 
and place constraints on the dynamical history of the observed system.

\begin{figure*}
\includegraphics[width=7.7cm,angle=-90]{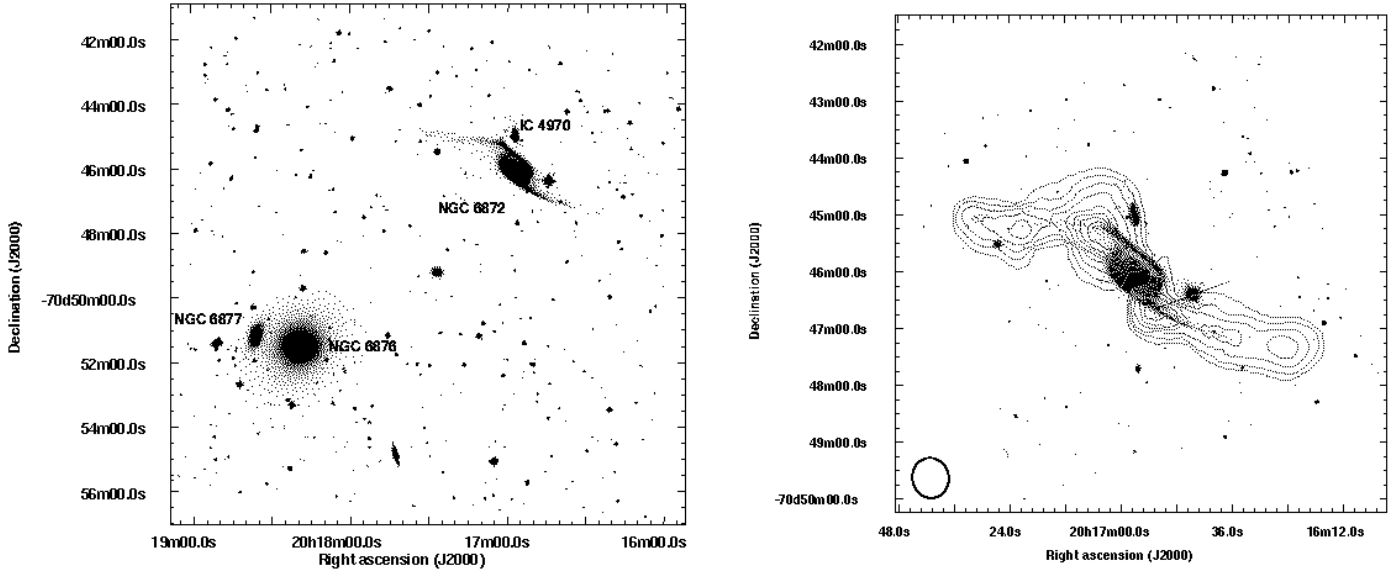}
\caption{
(a) Optical image from the Digitized Sky Survey of the field around 
NGC~6872. 
North is up and east is left. 
\cite{2005ApJ...630..280M} 
detected diffuse X-ray emission in the region between NGC~6872 and the massive
elliptical galaxy NGC~6876. 
(b) Integrated \hi map overlaid on the blue-band VLT image. 
The contour levels increase from 
$14.4\times 10^{19}$ cm$^{-2}$ 
to $115.4\times 10^{19}$ cm$^{-2}$
by steps of 
$14.4\times 10^{19}$ cm$^{-2}$. 
The synthesized beam is shown in the bottom left corner. 
The frame is 155~kpc on each side for the adopted distance. 
}
\label{fig1mom0}
\end{figure*}

\section{\hi Observations}
\subsection{Observations and data reduction} 
\hi observations were carried out using the ATCA 
in five different configurations. 
Details are given in Table~2. 
The correlator provided a total bandwidth of 16 MHz divided in
512 channels, which corresponds to a total velocity coverage of
3431 km~s$^{-1}$ and a velocity resolution of 6.7 km~s$^{-1}$. 
The data were reduced in {\sc miriad} and {\sc aips} using standard procedures. After
substantial flagging (narrow-band interferences around 1400, 1403 and
1406 MHz
plus broad-band solar interference, all affecting the shortest
baselines) the
data were calibrated using PKS 1934--638 which lies only 8\degr\ away
from
NGC~6872. The 20-cm continuum emission was subtracted by fitting a
first
order base level to all line-free channels. The four good \hi\ line
data sets
were then Fourier-transformed together using `robust' weighting
({\tt ROBUST}=0.5).
To enhance the signal-to-noise ratio of the \hi emission we smoothed the
data to
a velocity resolution of $\sim20$~km~s$^{-1}$ . All velocities are in the barycentric
reference frame using the optical definition.
The final maps obtained by combining the various data sets have 
a synthesised beam of
$42\farcs4
\times 38\farcs8$ and a noise level per channel of 0.90~mJy, 
which is very close to the theoretical value of
0.85~mJy.
Moment maps (zeroth order, first and second order) were obtained using
signals 
above 2 mJy\,beam$^{-1}$.

\hi masses are related to the \hi integrated intensities 
$\int S_{\rm HI} dv$ (in Jy~km~s$^{-1}$) by: 
\begin{equation}
M({\rm HI}) (M_\odot) = 2.36\times 10^5 D_{\rm Mpc}^2 \int S_{\rm HI} dv 
\end{equation}
where $D_{\rm Mpc}$ is the distance to the galaxy in Mpc. 

\hi column densities can be derived from the observed integrated intensities by
using the relation:
\begin{equation}
N_{\rm HI} = {{110.4\times10^3}\over{a b}} \int S_{\rm HI} dv 
\end{equation}
where 
$N_{\rm HI}$ is the column density of hydrogen atoms in 10$^{19}$~cm$^{-2}$ 
and $a$ and $b$
are the diameters of the synthesized beam at full width half maximum (FWHM) in arcseconds.

\begin{table}[t]
\begin{tabular}{ll}
\noalign{\smallskip}
\hline
\noalign{\smallskip}
Arrays,			&750D,		25/09/2001, 	12h\\
observing dates		&1.5D (corrupt), 17/11/2001,	$<$6h\\
and durations		&EW352, 	26/12/2001,	12h\\	
			&1.5B,	 	07/07/2002,	12h\\
			&750B, 		03/08/2002,	12h\\
Pointing position (J2000)	&$20^{\rm h}17^{\rm m}38^{\rm s}$; 
		$-70^\circ48'47''$\\

Primary beam		&$34\farcm1$\\
Central frequency	&1399 MHz\\
Number of channels 	&512\\
Calibrator		&PKS~1934--638 (14.91 Jy)\\ 
Combined data:	\\
Synthesized beam	\\
- Major $\times$ minor axis 
	&$42\farcs4\times38\farcs8$ (FWHM)\\
- Position angle	& $+12^\circ$\\
Channel width		&$\Delta v = 20.55$ km s$^{-1}$\\
Noise level ($1\sigma$)	&0.9 mJy beam$^{-1}$channel$^{-1}$\\
			&$1.24\times10^{19}$ cm$^{-2}$ channel$^{-1}$\\
			&$1.6\times10^7 M_\odot$ beam$^{-1}$channel$^{-1}$\\
\noalign{\smallskip}
\hline
\end{tabular}
\caption[]{Summary of \hi observations.}
\end{table}

\begin{figure*}
\includegraphics[width=12cm,angle=-90]{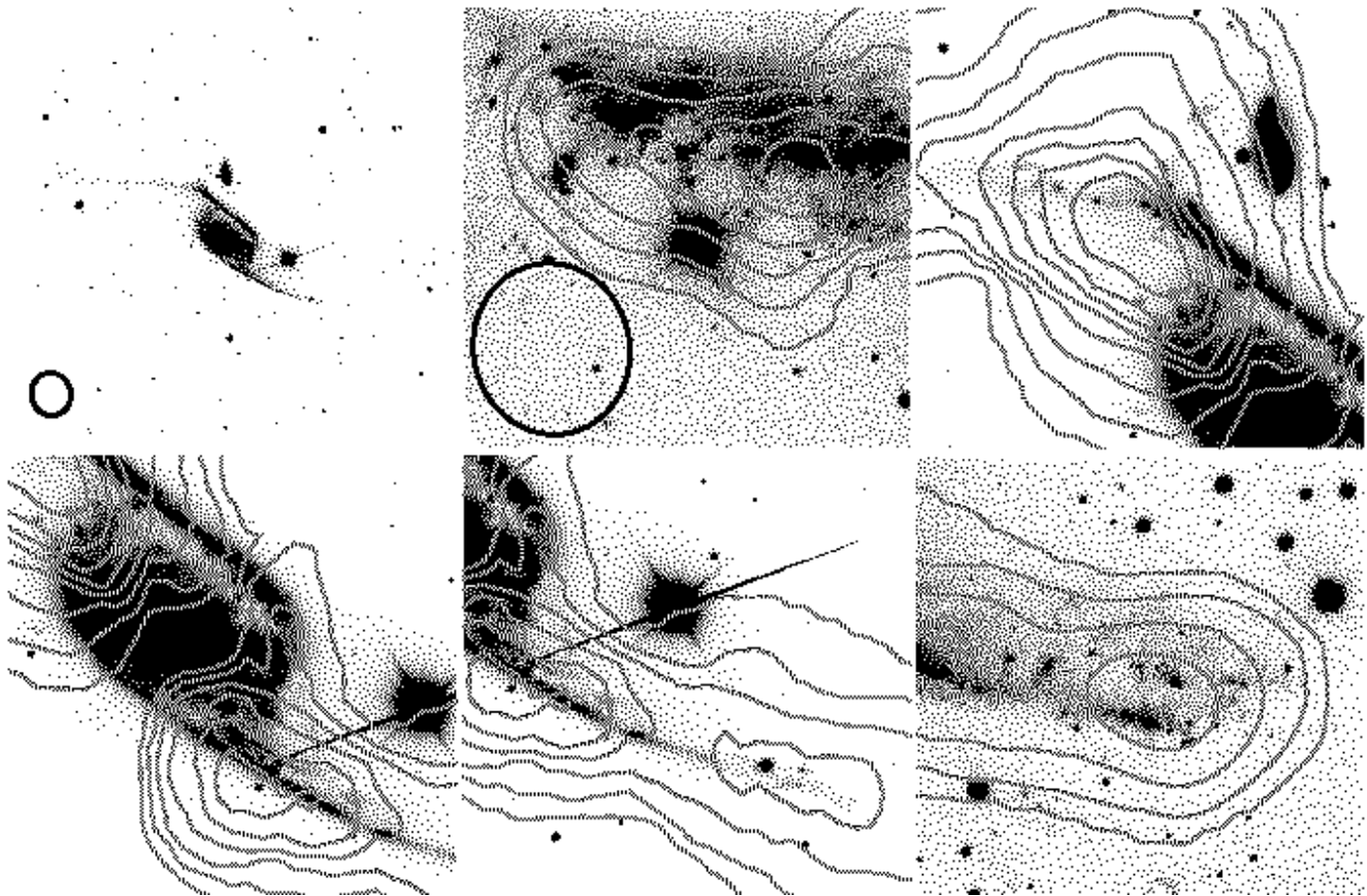}
\caption{Details of the \hi distribution as contours overlaid on the VLT B-band image, 
as in the previous figure. The size of the beam is indicated in the lower left corner
of the first two frames. Aside from the first frame, all the other frames have the
same size. 
North is up and east is left. 
The contour levels are the same as in the previous figure. 
In Figure~\ref{fig4hispec} the \hi concentrations in the southern tail are labeled S1 to S3 
and those in the northern tail N1 and N2. 
Their \hi content is given in Table~\ref{tabhiregions}. 
}
\label{fig2mom0details}
\end{figure*}
\begin{figure}
\includegraphics[width=8.0cm,angle=-90]{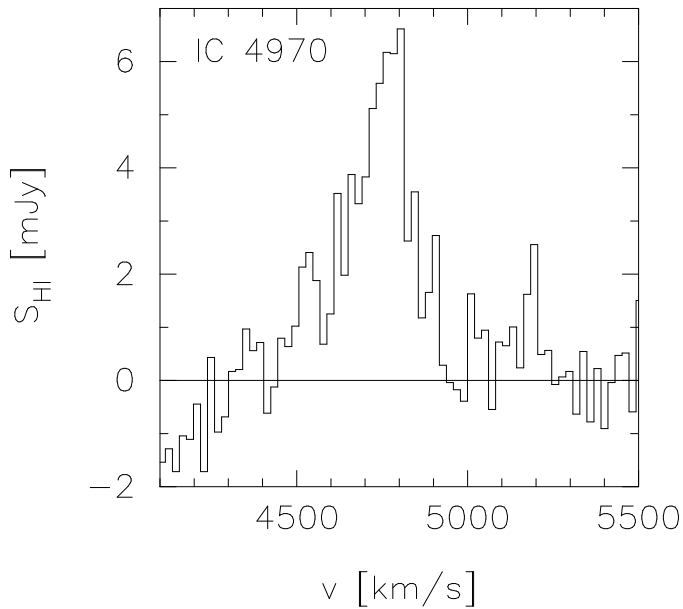}
\caption{\hi spectrum toward the companion IC~4970. }
\label{fig3hicompa}
\end{figure}

\subsection{Results}

\subsubsection{\hi distribution}

\begin{figure*}
\includegraphics[width=17.5cm]{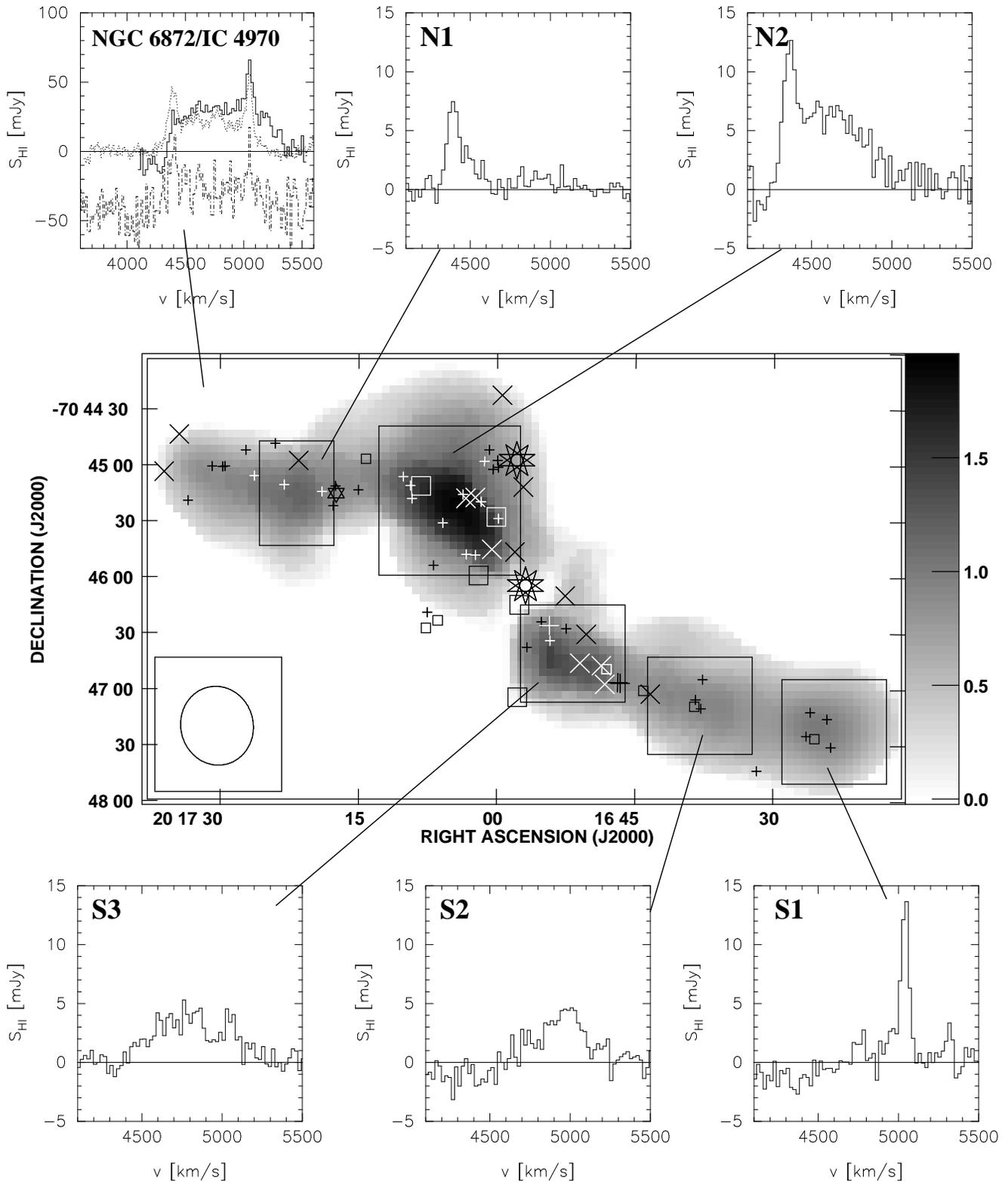}
\caption{Grey-scale map of the \hi distribution and \hi spectra measured over 
several regions in the tails labeled S1, S2 and S3 (southern concentrations) 
and N1 and N2 (northern concentrations).  
The first spectrum 
was measured in the rectangular area around 
the NGC~6872/IC~4970 system. 
The solid line refers to the ATCA observations, the dotted line is the single-dish 
Parkes spectrum of \cite{1997A&AS..126....3H} 
and the dotted-dashed line is the Parkes HIPASS spectrum, which has been 
shifted by $-40$~mJy for clarity. 
The ten-pointed stars mark the positions of the centers of NGC~6872
and of IC~4970. 
The location of the star clusters detected by \cite{2005A&A...435...65B} 
is indicated. 
Star clusters less massive than $10^6 M_\odot$ are marked by small symbols, and 
more massive clusters by large symbols. 
Plus signs and squares correspond to star clusters younger than 100~Myr, and
crosses and stars of David to older clusters. 
The squares and stars of David indicate a large extinction, with $A_v > 1$. 
Note that the regions S1 and S2 at the tip of the southern tail 
contain only young star clusters. 
}
\label{fig4hispec}
\end{figure*}

\begin{figure*}
\includegraphics[width=9.5cm,angle=-90]{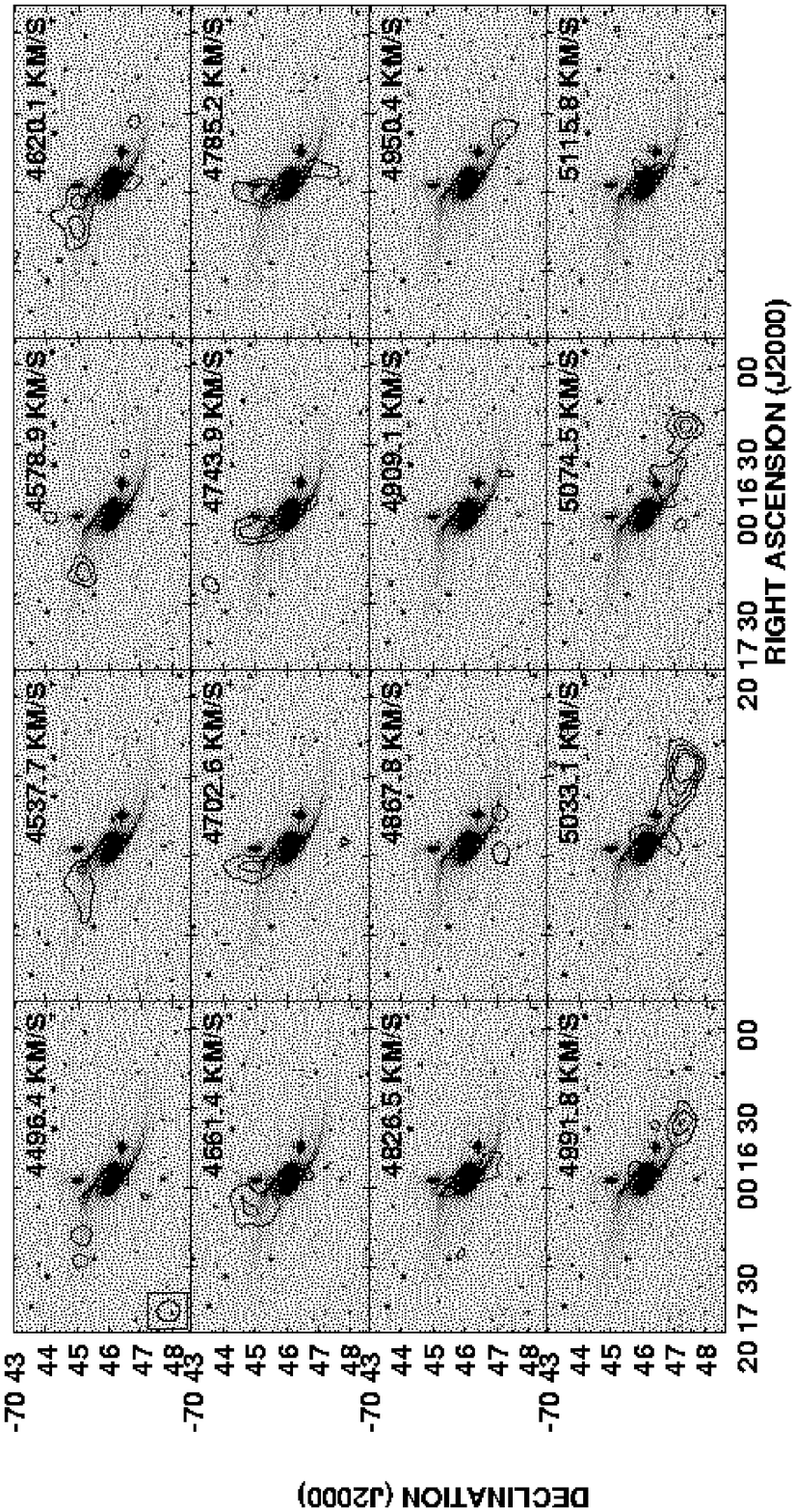}
\caption{The \hi emission of NGC~6872 overlaid on a grey scale optical image from the 
Digitized Sky Survey and displayed as channel maps running from 4496~km s$^{-1}$
(top left panel) to 5116~km s$^{-1}$ (bottom right panel). 
Every other channel is shown 
of the channel maps showing \hi emission. 
North is up and east is left. 
The levels are 3, 5, 7, 9 mJy/beam. 
The first contour corresponds to the $\sim3\sigma$ level. 
The synthesized beam is shown in the bottom left corner of the first panel. 
}
\label{fig5channels}
\end{figure*}

Figures~1b and \ref{fig2mom0details} 
display the \hi distribution as iso\-contours 
superimposed on the blue-band VLT image on a grey scale. 
The atomic gas is confined to the interacting galaxies
NGC~6872/IC~4970 and there is no evidence 
for the presence of extended \hi gas around
the galaxies or in the region between NGC~6872 and 
the massive elliptical NGC~6876. 
We measure an \hi flux of 17.5~Jy~km~s$^{-1}$ toward 
NGC~6872/IC~4970, which is less than 
the published single-dish measurements within the $15'$ beam of 
the Parkes antenna:   
$20.1\pm 0.5$~Jy~km~s$^{-1}$ \citep{1997A&AS..126....3H}; 
21.7~Jy~km~s$^{-1}$  (\citealt{2004MNRAS.350.1195M},  
entry HIPASS J2016-770 
in the HICAT catalog). 
Since the single-dish flux is around 21~Jy~km~s$^{-1}$, as much as 
$\sim 3.5$~~Jy~km~s$^{-1}$ could be distributed in an extended area 
outside the main galaxies and below the detection limit of our interferometric
observations. This corresponds to as much as $\sim 3\times 10^9 M_\odot$ of 
\hi gas, or 20\% the amount of \hi gas contained in the NGC~6872/IC~4970 system. 
We can also use the noise level in our maps to estimate the \hi flux  
in the area of the X-ray trail, which 
\cite{2005ApJ...630..280M} 
define as a 
$4\farcm35\times 5\farcm9$ rectangular
region between the NGC~6872 and the large elliptical NGC~6876. This 
area is about 50 times larger than our synthesized beam. 
From the noise level given in Table~1, we can estimate the
noise equivalent flux density in that area as $NEFD=\sqrt{50}\sigma \simeq 6.4$~mJy. 
The radial velocity difference between the two galaxies is $\Delta v\sim 800$~km~s$^{-1}$, 
which gives a flux $NEFD\times\Delta v = 5.1$~Jy~km~s$^{-1}$. 
This value is clearly too high since it exceeds the difference 
between the interferometric and the 
single-dish measurements. 

The central region of NGC~6872 is devoid of atomic gas. 
\hi was found neither in absorption
toward the central continuum source, nor in emission. 
This is not unusual in spiral galaxies, where most of the gas in the center
is often in molecular form. From observations of the CO(J:1-0) line toward
the center of NGC~6872,  
\cite{1997A&AS..126....3H} 
inferred a mass of molecular hydrogen $M(H_2)$ of 
$9.6\times 10^8 M_\odot$ within the central 45$''$. 

The companion galaxy, IC~4970, is clearly detected in \hi 
(Fig.~\ref{fig3hicompa}). 
It contains $\sim 1.3\times 10^9 M_\odot$ of atomic gas. 

Five large \hi concentrations are seen in the outer parts of NGC~6872, 
two in the northern tail and
three in the southern one. 
Figure~\ref{fig2mom0details} shows that the most northern concentration  
is spatially roughly coincident with
blue stellar clusters. This is also where 
peaks in the H$\alpha$ distribution were seen 
(\citealt{1993ApJ...418...82M}). 
A significant amount of \hi is also found near the break in the northern arm, 
close to the nearby companion IC~4970. 
In the southern tail, the three \hi peaks all occur in regions where 
star clusters are seen in the blue-band image. 

There is a clear overall asymmetry in the \hi\ distribution: 
we estimate that the north-eastern part contains 
about 1.4 times as much atomic gas as the south-western part. 
\cite{2005A&A...435...65B} 
estimated a star formation rate of $\sim 16.5$ M$_\odot$yr$^{-1}$ 
in the north-eastern tail and about half that amount in the western tail, using
an empirical relationship between the specific $U$-band luminosity and the
star formation rate measured by \cite{2000A&A...354..836L}. 
In the center, they inferred a star formation rate about five times lower than in 
the north-eastern tail. 

Finally, we show integrated \hi spectra measured in different regions
of the galaxy, as illustrated in Figure~\ref{fig4hispec}. 
The regions in the southern tail have been labeled S1 to S3, 
and those in the northern tail N1 and N2. 
The amount of gas contained in each region is given in Table~\ref{tabhiregions}. 
Gas at the tip of the southern tail in region S1 produces a narrow \hi
line (width $\sim100$ km~s$^{-1}$) centered around 5050~km~s$^{-1}$. 
Emission from the other regions produces broader spectra. 
In region N2, which lies near the break in the northern arm, 
there are clearly several velocity components. 
Region N1 near the tip of the northern arm 
also exhibits a narrow \hi line. 

The global \hi spectrum integrated over the whole NGC~6872/IC~4970 system 
is shown in Fig.~\ref{fig4hispec} (first spectrum, upper left). 
The line is very broad ($\sim 950$~km~s$^{-1}$ at the base), 
in agreement with the Parkes single-dish \hi 
spectra 
(\citealt{1997A&AS..126....3H}, 
\citealt{2004MNRAS.350.1195M}), 
also shown in Fig.~\ref{fig4hispec}. 
Since \cite{1997A&AS..126....3H} used the radio convention
$v_{\rm rad}/c=\Delta\nu/\nu$ for velocities, we converted 
the values to show the spectrum using the 
optical definition $v_{\rm opt}/c=\Delta\lambda/\lambda$ 
used in the present paper. 
The low-velocity peak around 4400~km/s seen in the single-dish \hi spectra 
is not seen in the ATCA map. 

Also shown in Fig.~\ref{fig4hispec} is the location of the
star clusters detected by \cite{2005A&A...435...65B}. 
We have divided the star clusters into different groups, 
depending on their age (more or less than 100~Myr), 
mass (more or less than $10^6 M_\odot$), 
and extinction (more or less than $A_v=1$), 
using the estimates of  \cite{2005A&A...435...65B} 
based on the three dimensional spectral energy fitting algorithm of 
\cite{2003A&A...397..473B}. 
Interestingly, clusters in the outer regions of the tails 
(S1, S2, N1, and the region at the very tip of the northern tail) are 
predominantly young and low-mass and have a low extinction 
(small plus signs). 
Some old, low extinction massive clusters (larges crosses) are found 
north of the northern tail at the border of the \hi distribution. 
Such old massive clusters are also found in the inner region 
(regions S3 and N3). 
Several young, low-mass, low-extinction clusters (small plus signs) 
are found near the companion, slightly to the east, and it is likely 
that their formation has been triggered in the interaction. 
Regions S3 and N1 contain large amounts of \hi gas, and 
a collection of star clusters. The young ones (plus signs and boxes) 
seem to lie at the periphery of the \hi concentrations. 
There is no systematic indication that clusters with higher extinction
(boxes) lie closer to the peak of the \hi concentration. 
Star clusters are expected to form from denser, molecular gas
which does not necessarily coincide with the more diffuse atomic
gas observed here. Also, some clusters may lie in the front part of 
the \hi concentration and not suffer from much extinction. 
\cite{2005A&A...435...65B} 
found an excellent correlation between the location of the youngest clusters 
($< 10$ Myr) and the distribution of H$\alpha$ emission in the map of 
\cite{1993ApJ...418...82M}. 

\begin{table}[t]
\begin{tabular}{llll}
\noalign{\smallskip}
\hline
\noalign{\smallskip}
Region	&$\int S_{\rm HI}dv$	& $M(HI)$	&$d$\\
	&Jy km s$^{-1}$		&M$_\odot$	&kpc\\
\hline
\noalign{\smallskip}
NGC~6872/IC~4970 &17.5		&$1.54\times10^{10}$\\
NGC~6872 	&16.0		&$1.41\times10^{10}$\\
IC~4970		&1.46	&$1.3\times 10^9$\hfill\\
N1		&1.25	&$1.1\times10^9$	&17\\
N2		&4.40	&$3.9\times10^9$	&39\\
S1		&2.18	&$1.9\times10^9$	&53\\
S2		&1.03	&$0.90\times10^9$	&34\\
S3		&0.86	&$0.75\times10^9$	&12\\
\hline
\end{tabular}
\caption[]{
\hi fluxes and corresponding \hi masses. 
The values for NGC~6872 were obtained by subtracting 
those measured toward the companion
from the values measured for the pair.
The third column lists the distance on the sky of the peak of the various \hi concentrations 
labeled in Fig.~\ref{fig4hispec} 
to the center of NGC~6872.} 
\label{tabhiregions}
\end{table}

\subsubsection{\hi kinematics}
We now examine the kinematical information that can be gained from 
the \hi maps. Figure~\ref{fig5channels} shows the velocity channel maps
as isovelocity contours superimposed on an optical image. 
Gas in the north-eastern tail has a lower velocity 
along the line-of-sight (an heliocentric velocity around 
4500~km~s$^{-1}$) than the south-western tail 
(around 5100~km~s$^{-1}$).
The line-of-sight velocities vary smoothly across the galaxy 
and are characteristic of those of a rotating disk.

\begin{figure}
\includegraphics[width=8.0cm]{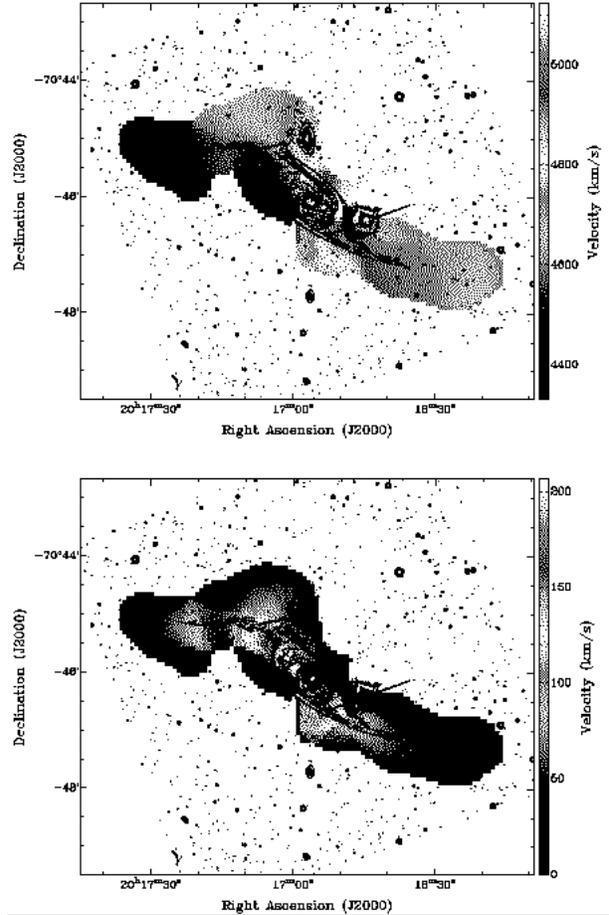}
\caption{Top: \hi velocity field in colors superimposed on the VLT blue-band image in contours.
Bottom: \hi velocity dispersion map in colors superimposed on the VLT blue-band image in contours.}
\label{fig6mom12}
\end{figure}

Figure~\ref{fig6mom12} shows 
the mean \hi velocity field (top)
and the \hi velocity dispersion (bottom). 
Again, the smooth velocity gradient across the galaxy appears clearly. 
The velocity dispersion map shows a peak on each side of the 
galaxy's center. One is located on the concave side near the break 
of the northern arm. The other one lies south of the main body of the 
galaxy where the southern tail begins. 
A region of increased velocity dispersion ($\sim75$~km~s$^{-1}$) 
coincides with the 
N1 \hi concentration in the northern tail.

\section{$N$-body simulations}

In order to narrow down the parameter space we 
first used a restricted three-body model.  
10~000 test particles were distributed in a 
Miyamoto potential representing a galactic disk. 
The particles were advanced according to the gravitational
effect of the potential of the host galaxy and that of a perturbing 
companion represented by a rigid Plummer potential moving on 
a Keplerian orbit. 
A good match was found for a prograde, parabolic, low-inclination
encounter with a companion five times less massive than the primary, 
in agreement with the findings of \cite{1993ApJ...418...82M}. 
Then we carried out self-consistent simulations with stars and gas, as
described below.  

\subsection{The model}

\begin{figure*}
\includegraphics[width=17.5cm]{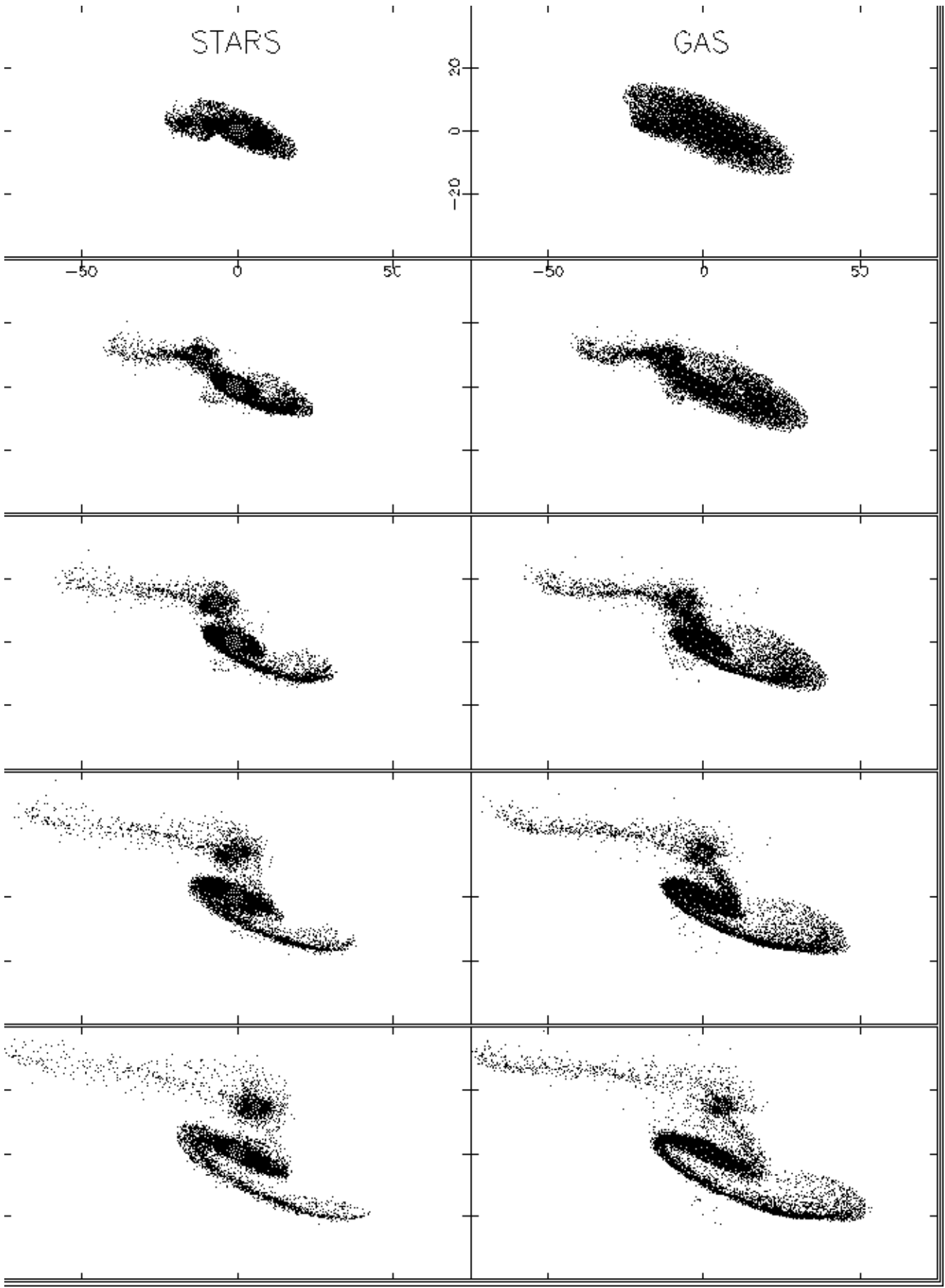}
\caption{Evolution of the stars (left column) and the gas (right column) 
in the simulation of NGC~6872/IC~4970. 
The stars in the bulge of the primary galaxy are shown in red. 
The system is viewed according to the observed geometry 
(position angle of 66$^\circ$, inclination of 73$^\circ$). 
The time step between two snapshots is 40~Myr, starting 
20~Myr after
perigalacticon. 
The position of the companion is indicated by a star. 
}
\label{fig7simul}
\end{figure*}

The simulations were performed using a particle-mesh algorithm ({\tt FFT3D}) that
makes use of the properties of Fast Fourier Transforms 
and of the convolution theorem. 
The gravitational potential is calculated at the nodes of a 
three-dimensional $128\times 128\times 64$ cartesian grid, 
providing a resolution of 
1~kpc.  
Effective use is made of the entire grid, 
rather than of only one eighth in standard Fourier algorithms, 
through implementation of a method described by
\cite{james77}. 

The main galaxy is made up of three components: a disk, a bulge,
and a halo. 
The halo is rigid and represented by a Plummer potential, 
$\Phi(r) = GM/(r^2 + a^2)^{1/2}$.
The bulge contains particles initially distributed in a Plummer potential.  
The disk contains two types of particles with different
velocity dispersions. 
The collisionless particles representing the stars are initially
distributed in a Toomre-Kuzmin disk so that the surface density has
the radial dependance 
$\Sigma(R) = M_G b /[2\pi(R^2 + b^2)^{3/2}]$. 
The vertical density profile corresponds to that of an 
isothermal sheet $\rho(z) \propto 1/ch^2(z/z_0)$ \citep{1942ApJ....95..329S}. 
The gaseous component is modeled as an 
ensemble of clouds, all of the same mass, that dissipate energy
through inelastic collisions with each other. 
Gas clouds approaching each other inside a cell 
collide and exchange energy every tenth time step 
(every 10$^7$ year). 
Initially, the gas follows a flat radial
distribution and accounts for about 10\%
of the mass of the stellar disk. The self-gravity of the gas is taken 
into account.  
Initially, the particles are assigned circular velocities 
in centrifugal equilibrium with the gravitational potential, 
and velocity dispersions are given by Toomre's stability criterion 
\citep{1964ApJ...139.1217T}. 

The companion galaxy is modeled as a rigid body represented by
a Plummer potential.
The asumption of a rigid structure for the companion is justified
insofar as we are interested in the effect of its passage on the
internal structure of the larger galaxy, which is probably 
little affected by the distortion of the companion. 

Setting the units of length and time to 1~kpc and 10$^7$ years, 
with the gravitational constant $G$ set to 1, gives a unit of 
mass of $2.22\times10^9 M_\odot$. The time step was set to 
10$^6$ years, and particles were advanced using the leap-frog
algorithm. 
We used 440~000 particles to model the
disk and the bulge of the primary galaxy 
(25\% of the particles being initially 
attributed to the bulge)
and 60~000 gas cloud particles.
The parameters of the simulation presented here are listed in 
Table~\ref{tabsimul}.

\subsection{Results}
\begin{table}[t]
\begin{tabular}{ll}
\noalign{\smallskip}
\hline
\noalign{\smallskip}
Orbit		& Parabolic \\
Mass ratio	& 5:1 \\
Main galaxy: 	& $M_{\rm disk}  = 120$, $a_{\rm disk} = 3$\\  
mass and scale length 	& $M_{\rm bulge} = 30$, $a_{\rm bulge} = 0.45$\\   
			& $M_{\rm halo}  = 90$, $a_{\rm halo} = 10$\\   
orientation	& $(15^\circ, 315^\circ)$ \\
Companion: mass and scale length & M$_c$ = 48, $a_c = 1$ 	\\
\noalign{\smallskip}
\hline
\end{tabular}
\caption[]{Parameters of the simulations. 
The orientation of the main galaxy is given by 
the angle between the orbital plane and the galactic plane ($15^\circ$) and 
the angle between the $x$-axis in the orbital plane and 
the projection of the vector normal to the main galaxy onto the orbital plane. 
 }
\label{tabsimul}
\end{table}

\begin{figure}
\includegraphics[width=8.cm]{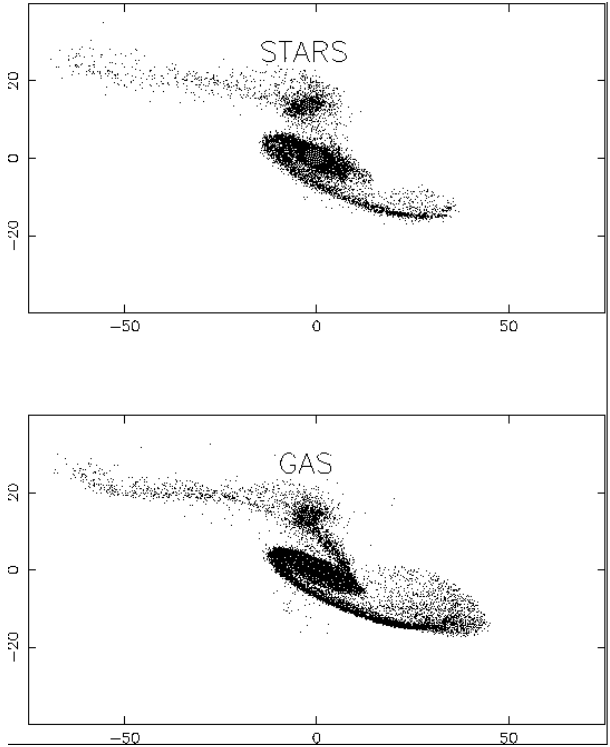}
\caption{Distribution the stars (top figure) and the gas (bottom figure) 
in the simulation 130~Myr after perigalacticon, 
when the simulated system resembles the observed one most. 
The position of the companion is indicated by a star. 
}
\label{fig8simulobs}
\end{figure}

\begin{figure}
\includegraphics[width=9.0cm]{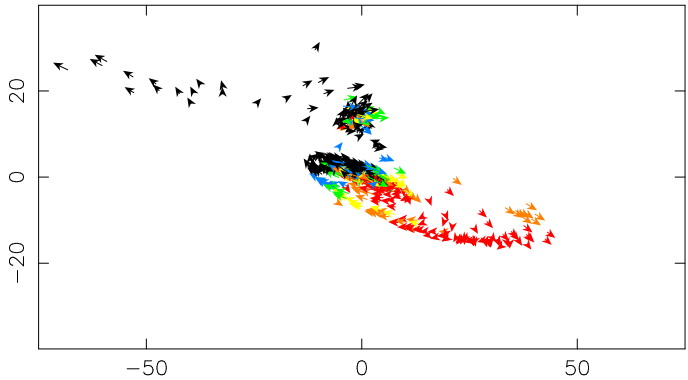}
\caption{Gas 
in the simulation 130~Myr after perigalacticon. 
The arrows indicate the direction of the velocity of the gas clouds projected 
on the plane of the sky. 
The colors indicate the velocity range along the line-of-sight, $v_{\rm los}$: 
larger than 200~km~s$^{-1}$ in red,
between 100 and 200~km~s$^{-1}$ in orange,
between 0 and 100~km~s$^{-1}$ in yellow,
between $-100$ and 100~km~s$^{-1}$ in green,
between $-200$ and $-100$~km~s$^{-1}$ in blue,
less than $-200$~km~s$^{-1}$ in black.
}
\label{fig9simulvel}
\end{figure}

\begin{figure}
\includegraphics[width=9.0cm]{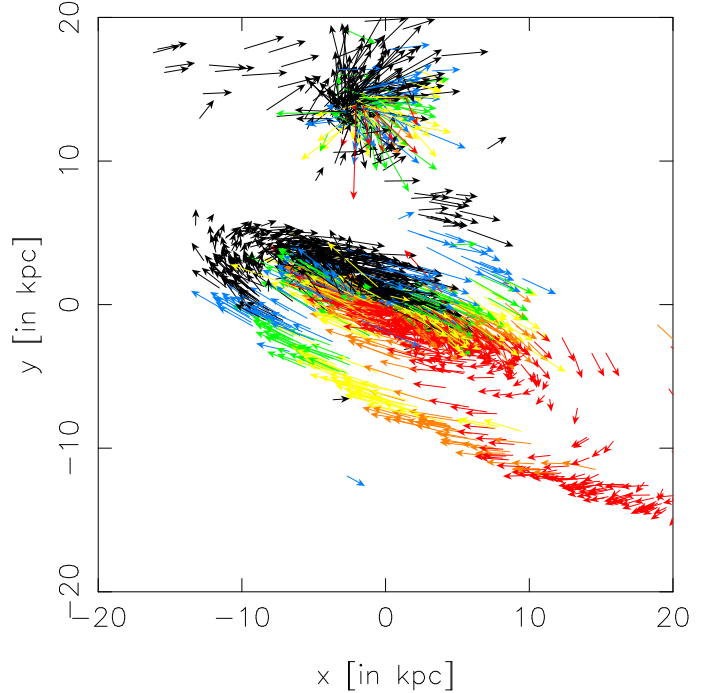}
\caption{Gas in the inner part of the simulated galaxy 
130~Myr after perigalacticon. 
The length of the arrows is proportional to the velocity on the
plane of the sky. 
The colors indicate the velocities along the line-of-sight, using  
the color code as in the previous figure. 
}
\label{fig10simulcentre}
\end{figure}

Figure~\ref{fig7simul} displays the evolution of the stars and the gas in the  
primary galaxy as it is perturbed by the companion passing on a prograde, parabolic
orbit.  
The gas distribution is initially more extended than that of the
stars. 
The time between two snapshots is 40~Myr, starting 20~Myr after closest approach. 
Despite its small mass (one fifth of the mass of the primary), the 
companion triggers prominent tidal tails very rapidly. 
The southern tail on the side of the galaxy opposite to the companion is 
thinner that the northern tail, where the perturbation is stronger. 
In the simulation the companion, modeled as a rigid Plummer potential
and displayed by a red star, accretes significant amounts of matter 
(both stars and gas). In the observed system, some diffuse stellar emission
is seen between the companion and the northern arm of the primary, but it is not clear 
that stars or gas have been dragged or have fallen onto the companion. 
Also, in the simulation the northern arm breaks at the location of the 
companion, whereas in the observed system both galaxies appear more clearly
separated. 
Aside from those differences, the model is able to reproduce several observed
features, including the extent of the tidal tails, the thinness of the southern tail, 
the central bar, the relative position of the two galaxies. The simulated
system resembles the observed one most around 130~Myr after closest approach, 
as shown in Fig.~\ref{fig8simulobs}.  
For that epoch, we also display the velocities of the gas in the simulations
(over the whole galaxy in Fig.~\ref{fig9simulvel} 
and in the inner part in Fig.~\ref{fig10simulcentre}). 
The arrows indicate the direction of the velocity of the gas on the plane of the
sky, whereas the colors are related to the velocities along the line-of-sight. 
The line-of-sight velocities are in good agreement with the observed velocity field. 
Interestingly, the simulation shows that gas in the outer part of the northern tail 
moves outward, whereas gas in the inner part moves inward toward the companion. 

The results of our simulations generally agree with those of \cite{1993ApJ...418...82M}. 
The gas is modeled in a similar way, as sticky particles, although  
in our simulation all gas clouds keep the same mass and do not evolve by 
coalescence or fragmentation. Star formation is not included in our model either. 
The initial conditions are slightly different: in our simulation, 
the orbital plane of the companion is slightly less inclined on the galactic
plane (15$^\circ$ instead of 25$^\circ$); 
their simulated halo is four times more massive than 
the stellar disk, whereas the spherical components of our model are
less massive (the added mass of the bulge and of the halo is the same as that
of the disk). This is why the tidal tails are slightly longer and thinner 
in our simulation. 
We find a blueshift of the companion 
relative to the main galaxy ($-192$~km~s$^{-1}$),  
as \cite{1993ApJ...418...82M} 
who found $-130$~km~s$^{-1}$.  
The true velocity difference between the two galaxies is uncertain since the published
observations cover a large range of values: 
For the companion, \cite{1988MNRAS.234.1051G} 
measured 4759~km~s$^{-1}$, while RC3 give 4727~km~s$^{-1}$. 
For NGC~6872, the following values are quoted: 
4626  \citep{1988MNRAS.234.1051G}, 
4688 (\citealt{2004MNRAS.350.1195M}, \hi spectrum), 
4717 (\citealt{1997A&AS..126....3H}, \hi spectrum), 
4818 (RC3); 
4892~km~s$^{-1}$ (\citealt{1997A&AS..126....3H}, CO spectrum). 
For comparison we converted the velocities quoted by \cite{1997A&AS..126....3H}
in the radio convention $\Delta\nu/\nu$ to the more commonly used optical 
convention $\Delta\lambda/\lambda$. 
The broad range of velocities across NGC~6872 (more than 900~km~s$^{-1}$ at the base
of the single-dish \hi spectrum) makes it difficult to 
accurately measure the systemic velocity. 
Taking the value from the CO spectrum of NGC~6872, which traces the molecular gas 
in the central 45$''$, and the RC3 value for the companion gives a 
blueshift for the companion of $-165$~km~s$^{-1}$, roughly consistent with the 
simulations. 

Within our model, the various parameters are well constrained by the characteristic
morphology of the observed system and we did not find a significantly different 
set of parameters
that would give an equally good match to the observations. 
This, however, could change with the addition of more parameters in the model 
(e.g. a halo spin or a different density profile, or an additional gravitational 
perturbation due, for instance, to a third galaxy). 

\section{Discussion}

\cite{2005ApJ...630..280M} 
discuss several possibilities to explain the diffuse X-ray emission  
seen between NGC~6872 and the large elliptical galaxy NGC~6876. 

-- 
As they argue, it is unlikely that the hot gas has been extracted from 
the elliptical galaxy  by {\it tidal interactions} because the
mass of the gas in the trail is more than three times larger than that 
contained in the elliptical. Had tidal interactions been at work, their
effect would have been seen also in the distribution of the stars in 
NGC~6876.  The isophotes of NGC~6876 are regular; 
Hubble Space Telescope observations do, however, reveal a slight central depression, 
which is not due to dust absorption. 
This has been interpreted as the possible signature of the end product of the merging of two gas-free stellar
system, each harboring a massive black hole 
\citep{2002AJ....124.1975L}. 
We can estimate the dynamical mass of NGC~6876 from the observed velocity
dispersion of 231~km~s$^{-1}$ \citep{1987ApJS...64..581D} %
using the relation 
$\left( \frac{M}{L} \right)_{\rm vir} = (5.0 \pm 0.1) \times R_e\sigma_e^2/(LG)$, 
which is the best-fitting virial relation found for a sample of 
25  elliptical and lenticular galaxies from the SAURON sample 
\citep{2006MNRAS.366.1126C}. 
Taking the effective radius $R_e = 25.48$~kpc as the half-light radius
in the blue-band given in NASA/IPAC Extragalactic Database, 
we obtain $M_{\rm vir} = 1.57\times 10^{12} M_\odot$. 
The projected distance on the sky between NGC~6872  and NGC~6876  is 
$\sim142$~kpc, and the radial velocity difference is about 800 km~s$^{-1}$. 
This gives a timescale of 180~Myr, 
assuming comparable contributions from the velocity in the plane of the sky 
and the separation along the line-of-sight.
If NGC~6872 had passed near NGC~6876,
a $\sim 10^{12} M_\odot$ galaxy around 180~Myr ago, it is likely that such an
interaction would have left a trace in the  \hi distribution, which is 
not observed.

-- The X-ray trail could be due to {\it ram pressure stripping} from NGC~6872 as the
spiral galaxy moves
supersonically though the hot gas within the Pavo group
\citep{2005ApJ...630..280M}. 
Ram pressure stripping is known to affect the \hi distribution in 
some galaxies (e.g. \citealt{2004A&A...419...35V}). 
However, we do not see any sign of that effect in the
observed \hi distribution of NGC~6872. 
Therefore, we cannot infer any information about the motion of NGC~6872 with respect to 
the surrounding gas in the Pavo group. 

-- Other mechanisms involving an {\it interaction between the interstellar medium
and the intergalactic medium} could be at work, such as turbulent viscous 
stripping \citep{1982MNRAS.198.1007N}, 
where gas at the interface between the galaxy and
the intergalactic medium could be heated. 
At the resolution of our \hi observations, we do not see any evidence for
that effect either. 

\begin{figure}
\includegraphics[width=8.0cm]{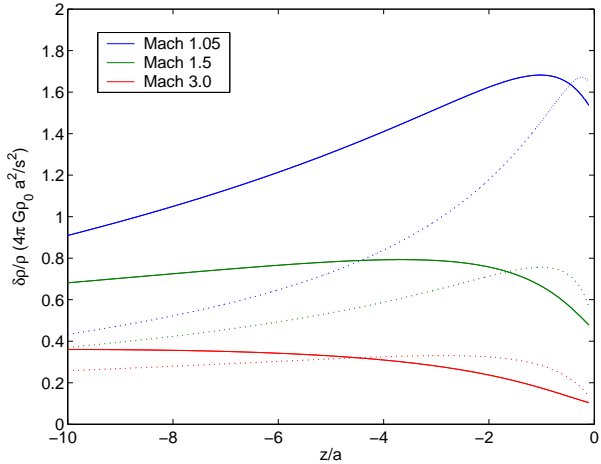}
\caption{
Density contrast $\delta\rho/\rho$ downstream of a 
Plummer potential  (solid lines) and an NFW potential (dotted lines) 
moving supersonically through a medium with different Mach
numbers. 
}
\label{fig11bondihoyle}
\end{figure}

-- {\it Accretion into a Bondi-Hoyle trail} is the most favored hypothesis. 
\cite{2005ApJ...630..280M} 
have calculated the extent of the trail assuming an NFW profile 
\cite{1996ApJ...462..563N} 
for the dark matter distribution in NGC~6872
with a concentration parameter $c=15$ and extending until a radius 
$ca$, where $a$ is the inner radius. 
An extended halo, with an inner radius $a\simeq 20$~kpc, 
is required to reproduce the observed
length of the trail. 
We have performed the same calculation, using a different profile for the halo, 
namely the one corresponding to the Plummer potential that we have
used in the simulation with a scale length of 10~kpc.  
As illustrated in Fig.~\ref{fig11bondihoyle}, for a Plummer profile 
the downstream profile of 
density contrast peaks at a larger distance from the moving halo, and 
decreases more softly. At a downstream distance $z=-10a$, the density contrast 
is slightly higher for the Plummer sphere than for the NFW profile. 
Although the density contrast along the trail is, in principle, sensitive
to the density profile of the moving halo, this is difficult to use in practice. 
This calculation is simplified, based on linearized flow equations, which 
applies only for small density perturbations and 
\cite{2005ApJ...630..280M} 
point out that the observed temperature ratio between the trail and the 
intra-cluster gas implies an overdensity larger than one. 

More sensitive X-ray observations would help elucidate
the nature of the X-ray trail. 
As discussed above, gravitational interaction with the large elliptical 
is unlikely to have produced such an X-ray trail without any \hi counterpart. 
More detailed simulations of 
the evolution of a spiral galaxy in the gravitational potential of 
a group, interacting with the hot intra-cluster medium and 
with both a massive elliptical and a smaller galaxy are required to 
shed more light on the relative importance of the different mechanisms. 
This is beyond the scope of this paper.  

\section{Summary and conclusion}

In this paper we have presented inter\-fe\-ro\-me\-tric \hi observations of the
interacting system NCG~6872/IC~4970 in the southern group of galaxies
Pavo obtained with the ATCA, and $N$-body numerical
simulations of the evolution of the system. 

\begin{enumerate}

\item 
We detected about $1.54\times 10^{10} M_\odot$ of \hi gas 
toward the NGC~6872/IC~4970 pair, out of which about 
$1.3\times 10^9 M_\odot$ is associated with the companion IC~4970. 
The gas in NGC~6872 is distributed in an extended rotating disk. 
Massive concentrations of gas ($\sim 10^9 M_\odot$)
are found at the tip of both tidal 
tails and towards the break seen in the optical northern arm near
the companion. 
An increased velocity dispersion is observed on each side of the
galaxy's center, near the break in the northern arm and south of the
main body where the southern tail starts; another region of higher
velocity dispersion coincides with the most northern \hi concentration. 
We have compared the \hi distribution with that of stellar clusters 
observed by \cite{2005A&A...435...65B}. 
Star clusters in the outer \hi concentrations in the tails 
are predominantly young and low-mass 
(less than 100~Myr and less than $10^6 M_\odot$) 
and have a low extinction
($A_v < 1$). 

\item No \hi gas was found associated with the X-ray trail 
observed by \cite{2005ApJ...630..280M}  
between NGC~6872 and NGC~6876, the dominant elliptical galaxy 
in the Pavo group located $\sim8'$ to the southeast. 
However, as much as $\sim 3\times 10^9 M_\odot$ of \hi gas could be 
hidden in extended regions of diffuse emission below our detection
limit. 
At the sensitivity and the resolution of the observations, 
there is no sign in the overall \hi distribution that 
NGC~6876 has affected the evolution of NGC~6872. 
There is no evidence either of ram pressure stripping. 

\item The simulations of a gravitational interaction with the small
nearby companion IC~4970 on a low-inclination prograde passage 
are able to reproduce most of the 
observed features of NGC~6872, including the 
general morphology of the galaxy, the inner bar, 
the extent of the tidal tails and 
the thinness of the southern tail. 

\item 
Our simulations do not model the gas environment within the group or the
motion of NGC~6872 through the group. Without those interactions 
we cannot address the hydrodynamic processes, i.e. ram pressure stripping, 
turbulent viscous stripping, or Bondi-Hoyle accretion, discussed in Section~4. 
However, together with the \hi maps, our simulations provide important constraints on viewing angle, internal velocities, 
mass and gas distributions for input into future hydrodynamic simulations. 

\end{enumerate}

\begin{acknowledgements}
We thank the referee for a very thoughtful report.
We are grateful to Fran{\c c}oise Combes for providing us a version of the code, 
to Vassilis Charmandaris for his help with the VLT data, 
and to John Black for useful comments on the manuscript. 
We thank E. de Blok, E. Sadler and E. Muller for their help
during the \hi\ observations. 
We thank Chris Mihos for providing us his H$\alpha$ data and Nate Bastian for 
stimulating discussions. 
This research has made use of the NASA/IPAC Extragalactic Database (NED) 
which is operated by the Jet Propulsion Laboratory, California Institute of Technology, 
under contract with the National Aeronautics and Space Administration.
The \hi\ observations were done with the Australia Telescope Compact Array, which is funded by the
Commonwealth of Australia for operations as a National Facility 
managed by the Commonwealth Scientific and Industrial Research Organisation 
(CSIRO). 
\end{acknowledgements}

\bibliography{6023-biblio}

\end{document}